# Local alterations of left arcuate fasciculus and transcallosal white matter microstructure in schizophrenia patients with medication-resistant auditory verbal hallucinations: A pilot study

Fanny **Thomas** [a,b*], Cécile **Gallea** [c], Virginie **Moulier** [a,d], Noomane **Bouaziz** [a], Antoni **Valero-Cabré** [b,e]† and Dominique **Januel** [a,f]†

[a] Centre de Recherche Clinique, Établissement Public de Santé de Ville-Evrard, 202 avenue Jean Jaurès, 93330 Neuilly-sur-Marne, France.
[b] Cerebral Dynamics, Plasticity and Rehabilitation Group, FRONTLAB, Centre de Recherche de l'Institut du Cerveau, CNRS UMR 7225, INSERM UMRS 1127 and Université Pierre et Marie Curie, 47 boulevard de l'Hôpital, 75013 Paris, France.
[c] Movement Investigations and Therapeutics, MOV'IT, Centre de Recherche de l'Institut du Cerveau, CNRS UMR 7225, INSERM UMRS 1127 and Université Pierre et Marie Curie, 47 boulevard de l'Hôpital, 75013 Paris, France.
[d] Centre Hospitalier du Rouvray, University Department of Psychiatry, 76301 Sotteville-lès-Rouen, France
[e] Laboratory for Cerebral Dynamics Plasticity and Rehabilitation, Boston University School of Medicine, 700 Albany Street, Boston, MA W-702A, USA.
[f] Université Sorbonne Paris Nord, Campus de Bobigny, 1 rue de Chablis, 93000 Bobigny.

† Prof. Antoni Valero-Cabré and Prof. Dominique Januel have equally contributed to the supervision of this work and share senior authorship.

**Correspondence**

***Fanny Thomas** PhD, EPS Ville-Evrard, URC, Pavillon Vendée, 202 avenue Jean Jaurès, 93330 Neuilly sur Marne, FRANCE. Email: fanny.thomas1@gmail.com

**Declarations of interest:** none

**ABSTRACT**

Auditory verbal hallucinations (AVH) in schizophrenia (SZ) have been associated with abnormalities of the left arcuate fasciculus and transcallosal white matter projections linking homologous language areas of both hemispheres. While most studies have used a whole-tract approach, here we focused on analyzing local alterations of the above-mentioned pathways in SZ patients suffering medication-resistant AVH.

Fractional anisotropy (FA) was estimated along the left arcuate fasciculus and interhemispheric projections of the rostral and caudal corpus callosum. Then, potential associations between white matter tracts and SZ symptoms were explored by correlating local site-by-site FA values and AVH severity estimated via the Auditory Hallucinations Rating Scale (AHRS).

Compared to a sample of healthy controls, SZ patients displayed lower FA values in the rostral portion of the left arcuate fasciculus, near the frontal operculum, and in the left and right lateral regions of the rostral portion of the transcallosal pathways. In contrast, SZ patients showed higher FA values than healthy controls in the medial portion of the latter transcallosal pathway and in the midsagittal section of the interhemispheric auditory pathway. Finally, significant correlations were found between local FA values in the left arcuate fasciculus and the severity of the AVH's attentional salience.

Contributing to the study of associations between local white matter alterations of language networks and SZ symptoms, our findings highlight local alterations of white matter integrity in these pathways linking language areas in SZ patients with AVH. We also hypothesize a link between the left arcuate fasciculus and the attentional capture of AVH.

**Keywords**: Schizophrenia, auditory verbal hallucinations, tractography, left arcuate fasciculus, transcallosal pathway.

# INTRODUCTION

Auditory verbal hallucinations (AVH) are linguistic percepts occurring in the absence of appropriate external language stimuli, a core symptom of schizophrenia spectrum disorders (SZ) (Andreasen and Flaum, 1991; Lim et al., 2016). Among other dysfunctions, prior studies have found AVH to be underpinned by alterations of white matter microstructure in the arcuate fasciculus (Shergill et al., 2007; Seok et al., 2007; Catani et al., 2011; de Weijer et al., 2011, 2013; Geoffroy et al., 2014; Leroux et al., 2015, 2017; McCarthy-Jones et al., 2015; Chawla et al., 2019; Falkenberg et al., 2020) and interhemispheric white matter pathways throughout the corpus callosum (Mulert et al., 2012; Steinmann et al., 2014; Leroux et al., 2015, 2017; Wigand et al., 2015). Nonetheless, these previous studies analysed rather long and complex tracts as single structures, despite them branching along their trajectory at multiple levels. These tractography methods do not take into account the influence of local rather than global microstructural abnormalities in on SZ and some of its symptoms.

The arcuate fasciculus is an association bundle linking the frontal lobe and the temporoparietal region (Catani and Thiebaut de Schotten, 2008). In the left hemisphere, it is the main connection between Broca's and Wernicke's areas, hence a key bundle of the perisylvian system in charge of the production and perception of speech (Paulesu et al., 1993; Vigneau et al., 2006; Price, 2010). In SZ patients, AVH have been associated with impaired connectivity in this language network and more specifically with a reduction of mean fractional anisotropy (FA), suggesting a loss of white matter integrity compared to healthy controls (Geoffroy et al., 2014). Nevertheless, an hyperconnectivity in language networks associated with AVH (Hubl et al., 2004; Rotarska-Jagiela et al., 2009; Abdul-Rahman et al., 2012) has also been reported (Thomas et al., 2016, for a review).

Among the few studies investigating the relationship between structural integrity of the left arcuate fasciculus and AVH severity in SZ (Rotarska-Jagiela et al., 2009; Ćurčić-Blake et al., 2015; Psomiades et al., 2016), two reports concluded a positive correlation between mean FA values in the left arcuate fasciculus and this symptom (Rotarska-Jagiela et al., 2009; Psomiades et al., 2016), whereas a third study yielded the opposite outcome (Ćurčić-Blake et al., 2015). Such heterogeneity could be explained by biases tied to a methodological approach in which white matter alterations were averaged over the whole tract and reduced to a single estimate, hence neglecting the influence of local structural complexity generated by local branching patterns.

Likewise, interhemispheric projections via the corpus callosum link homotopic areas of the left and the right hemisphere and their alterations have been also highlighted in SZ (Kubicki and Westin, 2002; Kyriakopoulos et al., 2008; Ellison-Wright and Bullmore, 2009) and associated to AVH (Mulert et al., 2012; Steinmann et al., 2014). More specifically, AVH have been linked to an increase structural callosal connectivity in patients with AVH compared to both healthy participants and to SZ patients without AVH and shown increased FA values in the genu of the corpus callosum connecting the left and right inferior frontal regions of speech (Hubl et al., 2004) and also in the splenium of this same structure connecting the left and right temporal auditory areas (Mulert et al., 2012). A study comparing SZ patients to healthy participants also reported lower FA values for the whole corpus callosum and for the caudal portion of the genu, and significant negative correlation between FA and AVH severity only for the latter area (Knöchel et al., 2012). Moreover, functional magnetic resonance imaging (fMRI)-based effective connectivity measures used to isolate directional influences between areas, showed significantly weakened connectivity from Wernicke's to Broca's areas, but also a trend toward a reduction in connectivity from contralateral regions of Broca's and Wernicke's areas to Broca's area in SZ patients with AVH compared to healthy participants (Curcic-Blake et al., 2013). The latter finding could support the hypothesis that interhemispheric fibres crossing the genu of the corpus callosum could be associated with AVH in these patients. The interhemispheric auditory pathway connecting bilateral auditory areas has also been involved in the pathogenesis of AVH and shown decreased FA values in SZ patients with AVH compared to patients without AVH and healthy controls (Wigand et al., 2015; Leroux et al., 2017). Opposite results have also been reported in first-episode SZ patients (Mulert et al., 2012) and in acute SZ patients (Hubl et al., 2004). Taken together, these findings suggest structural connectivity alterations in the interhemispheric auditory pathway in patients with AVH, which would depend on the age or the stage of the illness. A more advanced age and a longer period of illness have been associated with weakened white matter connectivity affecting the interhemispheric auditory pathway in patients with AVH.

Prior tractography reports in SZ employed whole-tract standard tractography methods that reduced tract complexity to a single mean FA measure per participant. This approach has the advantage of reducing the risk of false discoveries caused by multiple comparisons. Nonetheless, the influence of relevant local tract features particularly in key branching loci (i.e., sites of incoming and/or exiting fibres) and their role subtending AVH has the risk to pass unnoticed. To overcome this limitation, more recent alternative approaches investigate structural variation along one or more fasciculi of interest in intact (O'Donnell et al., 2009; Colby et al., 2012) and altered white matter connectivity (Concha et al., 2010; Yeatman et al., 2012).

Inspired by such precedent, we used an along-tract method for tractography, sensitive to local white matter features (Colby et al., 2012), to identify potential local differences of white matter integrity along the left arcuate fasciculus and the corpus callosum in SZ patients with medication-resistant AVH compared to a group of healthy participants.

## METHODS

**Participants**

The study included 17 patients diagnosed with SZ (according to the Diagnostic and Statistical Manual of Mental Disorder 5th edition, DSM-5) and suffering medication-resistant AVH, defined as unresponsive to treatment with at least two different antipsychotic drugs (one of them atypical) for at least 6 weeks. Patients showed clinical stability for at least 3 months (i.e., no changes in symptomatology, nor in antipsychotic medication during 3 months prior to the inclusion in the study). The study was approved by a local ethics committee (Ile-de-France III, Tarnier-Cochin hospital, France, ID-RCB 2014-A01595-42). All participants received a clear and detailed information about the protocol and provided written consent to participate. The presence and the severity of AVH were assessed with the Auditory Hallucinations Rating Scale (AHRS) across seven items investigating frequency, level of reality, loudness, number of voices, length of the content, level of distraction, and distress (Hoffman et al., 2003). The overall severity of SZ symptomatology was evaluated using the Positive And Negative Syndrome Scale (PANSS; Kay et al., 1987).

MRI anonymized data from 22 healthy controls obtained with the same MR device and sequences were used for comparison (Méneret et al., 2017; Welniarz et al., 2017, 2019). Patients and healthy subjects did not differ significantly regarding age or sex (**Table 1**).

**MRI acquisition**

MRI data were obtained using a Siemens VERIO 3 Tesla scanner (Siemens Healthcare, Erlangen, Germany) with a 32-channel head coil. Three-dimensional, high-resolution, isovoxel, T1-weighted brain volumes were recorded ($256 \times 240 \times 176$ matrix size with 176 contiguous slices, field of view (FOV) = 256 mm, 1mm isotropic resolution, sagittal slice orientation, repetition time (TR) = 2300 ms, and echo time (TE) = 2.98 ms). In addition, diffusion weighted images (DWIs) were acquired using a DWI sequence (104 x 104 x 84 matrix size with 60 contiguous slices, FOV = 208 mm, 2mm isotropic resolution, transversal slice orientation, flip angle = 90°, TR = 3800 ms, TE = 86 ms). The encoding protocol included 60 different non-collinear directions (gradient factor b = 1500 s/mm2) and 6 images without diffusion weighting used as the reference volume (b = 0 s/mm², b0 image).

**Region of interest (ROI) definitions**

Six ROIs were created to reconstruct the left arcuate fasciculus and the transcallosal fibres of the corpus callosum: Wernicke's area, the left and right inferior frontal gyrus, the left and right Heschl's gyrus/planum temporale and the corpus callosum. The left arcuate fasciculus included the Wernicke's area, defined as seed ROI, and the left inferior frontal gyrus (also known as Broca's area), defined as target ROI. The rostral transcallosal tract was reconstructed from the left inferior frontal gyrus (seed) to the right inferior frontal gyrus (target) through the corpus callosum. The interhemispheric auditory pathway corresponded to the reconstructed streamlines between the left (seed) and right (target) Heschl's gyrus/planum temporale and passing through the corpus callosum (Steinmann et al., 2014, 2019). The ROI corresponding to the Wernicke's area was a sphere of a 6 mm radius, centered at $x = −55$, $y = −41$, $z = 11$ and generated using the MarsBaR toolbox (http://marsbar.sourceforge.net/) (Brett et al., 2002). Please note that this ROI was the cortical site targeted with repetitive Transcranial Magnetic Stimulation (rTMS) to modulate AVH in the context of an ongoing study (see Thomas et al., 2019, for further details). The mask of the corpus callosum was manually drawn on a T1-weighted MNI template using the Functional Magnetic Resonance Imaging of the Brain (FMRIB) Software Library (FSL 5.0.9, http://www.fmrib.ox.ac.uk/fsl/). The binary masks of the left and right frontal inferior gyri and of the left and right Heschl's gyrus/planum temporale were extracted from the Harvard-Oxford Atlas (part of FSL software package) with a probability threshold of 30% (i.e., at least 30% probability that a given voxel is within frontal inferior gyri or Heschl's gyrus/planum temporale). These ROIs were de-normalized from the MNI space to the individual space using the inverse transformation generated during the segmentation of structural MRI volumes. Then, ROIs were resliced to DWI space for each subjects using do_fsl_reslice function available on Github and usable with Matlab software (https://github.com/romainVala/matvol/blob/master/Wrappers/do_fsl_reslice.m).

**Tractography analysis**

Diffusion volumes were first preprocessed using FSL software and then processed for probabilistic diffusion tractography with MRtrix3 software (http://www.mrtrix.org/). Diffusion volumes were corrected for motion and geometric distortions induced by eddy currents. The constrained spherical deconvolution method was used to estimate the fibre Orientation Distribution Function (ODF) in MRtrix3 (Tournier et al., 2019). Using a voxel-wise model of diffusion (the Q-ball model), the maximum-likelihood solution for fibre orientation within each voxel was represented by an ODF on the location of the fibre trajectory. To infer connectivity of crossing fibres, ODF information obtained from constrained spherical deconvolution was used with a probabilistic streamline algorithm with the entire ODF as a probability density function (ODF threshold = 0.1; step size = 0.5 mm x voxel size; radius of curvature = 1 mm; up-sampling of DWIs, data to 1 mm). Anatomically constrained tractography (Smith et al., 2012) was used to generate probabilistic streamlines of the three bundles of interest with a maximum path length of 180 mm, step size of 1 mm and back-tracking.

In the native space of each subject, a seed-to-target analysis was performed to reconstruct the left arcuate fasciculus and the transcallosal fibres of the corpus callosum. A reconstruction of white matter tract was considered to have failed if less than 10 streamlines were generated. Once reconstruction of these two tracts was completed, we verified its quality by visual inspection. When necessary, hand-drawn exclusion ROIs were created to remove spurious streamlines (i.e., isolated streamlines taking odd directions without the presence of actual underlying fibers). Thereafter, we estimated and averaged whole-tract FA values for each individual patient and healthy controls included in the study. A whole tract is defined as all streamlines reconstructed between a seed and a target area (insets A of Figures 1 and 2). Then, along-tract measures of FA were obtained on the basis of a B-spline resampling of fibres and the averaging of FA values for each individual fibre and tract location, defined by an elastic model of 42 spatial points located at analogous anatomical locations on each participant (Colby et al., 2012). For the left arcuate fasciculus, mean FA values were calculated for each tract point along mean tract fibres (Figure 1B) in the 'y' axis to explore local differences at specific tract points. For the rostral and caudal transcallosal tracts, mean FA values were determined for each tract point along mean tracts (Figure 2B and 3B, respectively) in the 'x' axis from the left to the right hemisphere. The number of points along two selected tracts, the arcuate fasciculus and rostral transcallosal projections, was established based on the average length of each tract (average length of the tract divided by 2). Although rostral and caudal transcallosal projections are longer than the arcuate fasciculus, 42 points were chosen for both bundles in order apply the same correction for multiple comparisons.

**Statistical analysis**

Data were analyzed with R software (R Core Team, 2017). Whole tract-averaged FA group values were compared between SZ patients and healthy participants using an analysis of covariance (ANCOVA) with age and sex as covariates. The equality of variances (Levene's test) and the normal distribution of the data (Shapiro-Wilk's test) were ascertained beforehand. A linear mixed-effects model was applied serially to each tract, whereas permutation tests served to adjust the *p*-values and control for type 1 errors (Colby et al., 2012). Age and gender were considered as nuisance covariables in these models. An outlier detection analysis on FA values was carried out for whole tract and along-tract FA values, separately by converting FA to z-scores. A threshold of +3 or -3 times the standard deviation was defined to identify outlying values. This analysis revealed a potential outlier in the SZ patient group for FA along the left arcuate fasciculus which was preserved as they were not false but simply atypical compared to other participants. No outliers were detected for the rostral transcallosal tract. Regarding the interhemispheric auditory pathway, the FA values of two patients were considered outliers for FA measures along the tract and were therefore excluded from the analyses.

Pearson correlations were performed between AHRS scores and FA values in each point and for each of the two white matter tracts. Then, permutation tests (5000 permutations) were performed to adjust *p*-values and control for multiple comparisons. Additionally, Pearson correlations (adjusted with Bonferroni correction) were performed between clinical scores (PANSS, AHRS and sub-items scores of AHRS) and FA values in areas in both tracts with significant differences between patients and controls.

## RESULTS

**Left arcuate fasciculus**

The whole-tract volume-weighted mean FA value for the left arcuate fasciculus did not differ significantly between SZ patients and healthy controls (mean FA values = 0.461 (SD = 0.026) vs 0.474 (0.021); $F(3, 35) = 1.03$, $p = 0.392$). In contrast, a mixed effects model revealed significant FA differences along this tract between the two groups. Intergroup differences, revealing lower FA values in patients were observed at sites number 31 to 34 (point 31, $p = 0.033$; point 32, $p = 0.008$; point 33, $p = 0.006$; point 34, $p = 0.022$), corresponding to the rostral portion of the arcuate fasciculus prior its ventral bending to reach Broca's area (**Figure 1A-B** and **1C**).

**Rostral portion of the transcallosal tract**

Whole volume-weighted mean FA values for the anterior transcallosal tract did not significantly differ between SZ patients and healthy controls (mean FA = 0.473 (0.020) vs 0.480 (0.020); $F(3, 35) = 0.85$, $p = 0.478$). Nonetheless, an analysis of tract local features revealed differences between the two groups at site number 8 to 10, 18, 25 to 27 and 34 to 36 (point 8, $p = 0.002$; point 9, $p < 0.001$; point 10, $p = 0.016$; point 18, $p = 0.014$; point 25, $p = 0.029$; point 26, $p = 0.043$; point 27, $p = 0.049$; point 34, $p < 0.001$; point 35, $p < 0.001$; point 36, $p = 0.001$), corresponding to the lateral and medial portions of the tract (**Figure 2A-B**). Specifically, for the lateral portions of the tract (site number 8 to 10 in the left hemisphere and sites number 34 to 36 in the right hemisphere), patients displayed significantly lower FA values than healthy controls.

In contrast, for the medial portions (site number 18 and 25 to 27), FA values were significantly higher in SZ patients compared to the healthy controls (**Figure 2C**).

**Interhemispheric auditory pathway**

The interhemispheric auditory pathway reconstruction failed for one patient. This subject was excluded from the statistical analysis. The whole-tract volume-weighted mean FA value did not differ significantly between SZ patients (n = 14) and healthy controls (mean FA = 0.554 (0.019) vs 0.545 (0.023); $F(3, 34) = 0.46$, $p = 0.715$). Statistical analysis along this tract showed higher FA values in patients compared to healthy controls at site number 17 to 23 (point 17, $p = 0.021$; point 18, $p = 0.010$; point 19, $p = 0.032$; point 20, $p = 0.035$; point 21, $p = 0.001$; point 22, $p = 0.001$; point 23, $p = 0.013$), corresponding to the medial section in the splenium of the corpus callosum (**Figure 3A-B and 3C**).

**Correlations with clinical symptomatology**

None of the significant correlations between FA of the left arcuate fasciculus, or transcallosal white matter projections, and AHRS scores survived correction for multiple comparisons (**Table 2**). To further explore a possible association between AVH and white matter integrity in our two fasciculi of interest, we correlated several SZ clinical scores with FA values on bundle sites for which a significant difference was found between SZ patients and healthy controls.

For the left arcuate fasciculus, our analysis reported no significant correlations of local FA values (at sites 31-34) with the PANSS score and with the AHRS total score after Bonferroni correction. However, significant positive correlations were found between 'attentional salience' and 'distress' sub-item scores of the AHRS and local FA values after correction for multiple comparison (**Table 3**). No significant correlation was reported for the other sub-items of the AHRS.

Regarding the rostral portion of the transcallosal tract and the interhemispheric auditory pathway, no significant correlation was found between local FA values and clinical scores (PANSS, AHRS total and sub-items scores) (**Table 3**).

## DISCUSSION

We investigated specific whole-tract and local alterations of white matter integrity for the left arcuate fasciculus and for interhemispheric pathways through the corpus callosum in SZ patients suffering medication-resistant AVH, compared to healthy controls. While whole-tract volume-weighted mean FA values were equivalent between the two groups, our findings showed that SZ patients with AVH had focal abnormalities of the left arcuate fasciculus, the interhemispheric transcallosal pathway linking the Broca's area with its homotopic right hemisphere region and the interhemispheric auditory pathway through the posterior part of the corpus callosum.

**The left arcuate fasciculus**

Our study supports the presence of fronto-temporal dysconnectivity in the left arcuate fasciculus, a tract known to play a critical role in AVH in SZ (Hubl et al., 2004; de Weijer et al., 2011; Abdul-Rahman et al., 2012; Geoffroy et al., 2014). This structural alteration was focal and located in the close vicinity of the left frontal operculum, which contains Broca's area. Our findings suggest that white matter abnormalities suffered by SZ patients do not necessarily affect the entire tract but only certain portions of this bundle. The arcuate fasciculus can be divided into two (anterior and posterior) superficial indirect short tracts and one deep direct long tract. The direct segment connects the fronto-temporal areas in the language network (Catani et al., 2005). Our study in SZ patients with AVH revealed alterations of white matter integrity in the rostral portion of the direct segment of the left arcuate fasciculus.

Anatomical variations of this direct segment were reported in the literature describing the existence of the frontal terminations in the dorsal premotor cortex in some healthy individuals, and not only in the inferior frontal gyrus (Frey et al., 2008; Bernard et al., 2019). To this regard, Abdul-Rahman et al. (2012) reconstructed the arcuate fasciculus, passing by the coronal plane by cutting the posterior limb of the internal capsule (at the precentral gyrus level in the frontal lobe) and the axial plane lateral to the sagittal stratum (at the superior temporal gyrus level in the temporal lobe) in SZ patients with (non-resistant) hallucinations compared to healthy controls. After discretizing the tract into 50 equally spaced points (and applying a parametrization method different than the one we applied), FA values revealed specific local abnormalities in the superior frontal segment of the left arcuate fasciculus, close to the left premotor cortex and the supplementary motor area. Importantly, these local abnormalities appeared when there were no significant differences in mean FA in the whole tract. Additionally, no significant correlations were found between FA values in the whole arcuate fasciculus and the PANSS score (total, positive, negative, or general subscale). By using a different delineation of the long direct segment of the arcuate fasciculus, our study confirms and extends prior findings by Abdul-Rahman et al. and contributes additional evidence regarding alterations of white matter integrity of the arcuate fasciculus in SZ with resistant AVH. Taken together, these findings suggest structural differences between SZ patients and healthy controls along the frontal aslant tract, a bundle involved in speech and language processing (Dick et al., 2019) that interconnects the superior (pre-supplementary motor area and precentral gyrus) and inferior (the pars opercularis/triangularis) frontal gyri (Catani et al., 2012). A previous study did not show structural differences in this tract between SZ patients and healthy subjects, but reported in patients a negative correlation between positive PANSS scores and mean diffusivity values in the frontal aslant tract, suggesting that those with higher positive symptoms scores (including hallucinations) may show higher white matter integrity (Leroux et al., 2018).

According to prior studies, lower local FA in SZ patients with AVH could reflect axonal damage such as decreased number, density or diameter of the axonal fibres, demyelination, loss of axonal membrane or loss of coherence (Beaulieu, 2002; Alba-Ferrara and de Erausquin, 2013) in the rostral portion of the direct segment of the left arcuate fasciculus close to the Broca's area, which is involved in the production of inner speech. This interpretation is consistent with a previous meta-analysis suggesting that AVH is caused by impairments affecting the production of inner speech (Kühn and Gallinat, 2012). Otherwise, our findings concur with reports showing a decrease in frontotemporal functional connectivity in the left hemisphere between the dorsolateral prefrontal cortex and superior temporal gyrus in SZ patients with AVH (Lawrie et al., 2002; Leroux et al., 2013, 2014), and suggest the possibility that reduced white matter integrity in the left arcuate fasciculus might cause functional dysconnectivity in the language network in SZ.

**The rostral interhemispheric fibres through the corpus callosum**

The comparison of SZ patients with AVH with healthy controls also revealed microstructural differences in interhemispheric transcallosal white matter pathways. More specifically, rostral fibres passing through the genu of the corpus callosum showed lower and higher FA values in lateral and medial portions, respectively.

Alterations of the structural integrity of large divisions of the corpus callosum in SZ patients have been reported for the last two decades. In such context, our study is the first to specifically investigate changes of local structural connectivity along the interhemispheric callosal fibres using site-by-site along-tract tractography analysis. Nonetheless, discrepancies persist with regards to the direction of such alterations in specific anatomical divisions or depending on the groups used for comparison. For example, several studies found lower callosal volume (Knöchel et al., 2012; Madigand et al., 2019) and lower FA values including the genu (Rotarska-Jagiela et al., 2008, 2009; Knöchel et al., 2012; Mulert et al., 2012; Wigand et al., 2015) in SZ patients compared to healthy controls. However, in SZ patients with AVH compared this time to non-hallucinating SZ patients, another report yielded FA increases in the rostral portion of the corpus callosum (Hubl et al., 2004).

According to such evidence, complex microstructural white matter callosal alterations, combining hyperconnectivity of interhemispheric pathways in the rostral corpus callosum with a hypoconnectivity in the vicinity of the left hemisphere Broca's area, and in homotopic right hemisphere regions should be further explored as potential hallmarks of SZ patients with AVH.

The homotopic region of Broca's area in the right hemisphere has been hypothesized to play a critical role in processing the emotional content of speech, resolving conflicting semantic information (for a review, see Price, 2010) and managing the transformation of speech segments, or phonemes, during the production of language (Kellmeyer et al., 2019). Moreover, rostral transcallosal fibres connecting Broca's area and its right hemisphere's homotopic region are important for the integration of different levels of linguistic complexity (Kellmeyer et al., 2019). On this basis, we hypothesize that in AVH, a hypoconnectivity between Broca's area and its contralateral right homotopic region would impact interhemispheric transfer and prevent adequate speech monitoring (Allen et al., 2008), hence leading to erroneous interpretations of emotional speech (Curcic-Blake et al., 2013).

Alternatively, reduced connectivity between inferior frontal gyri could suppress interhemispheric inhibitory interactions in the language network (Thiel et al., 2006) and result in their spontaneous activation during AVH. This explanation is supported by neuroimaging studies reporting unilateral left (Jardri et al., 2011), right (Copolov et al., 2003) or bilateral activations (Sommer et al., 2008) of the inferior frontal gyri during the generation of AVH in SZ. According to this explanatory model, AVH would be the result of insufficient inhibition of the right inferior frontal gyrus by the left hemisphere, and unexpected ectopic language production by the right hemisphere (Sommer et al., 2008; Sommer and Diederen, 2009). Additional investigations will be necessary to further understand the potential role of the rostral interhemispheric pathway in SZ with or without AVH, which at this point remains still speculative.

Regarding the increase of FA values in the medial portions of the rostral transcallosal tract, it might reflect an increase in information transfer between the left and right cerebral hemispheres. This increase could be caused by compensatory processes counteracting the hypoconnectivity observed in the lateral parts of this same tract and maintain the sharing and integration of information between both hemispheres. However, such interpretation of the functional role of the transcallosal fibres remains hypothetical <u>and at this point purely speculative</u>.

**The interhemispheric auditory pathway**

Our results showed higher FA values in the midsagittal section of the interhemispheric auditory pathway in SZ patients with resistant AVH compared to healthy controls. These findings differ from previous studies suggesting a loss of white matter integrity in this bundle in chronic SZ patients with AVH (Wigand et al., 2015; Leroux et al., 2017). Nevertheless, studies investigated brain structural connectivity of the interhemispheric auditory pathway in SZ suggested a progressive deterioration of the white matter fibres with age or duration of illness (Steinmann et al., 2019). SZ patients with AVH and with a short duration of illness (5-7 years) had increased FA values compared to patients without AVH or healthy controls (Hubl et al., 2004; Mulert et al., 2012), while patients with longer period of disease (11-17 years) showed a reduction of structural connectivity in the interhemispheric auditory pathway (Wigand et al., 2015; Leroux et al., 2017). Considering that SZ is a neurodevelopmental disorder showing brain alterations at different stages of illness (Weinberger, 1987), these findings suggest a gradient evolution of the integrity of white matter fibres in the course of disease or the age. We suggest that our patient sample might be located in the continuum of this gradient. However, our hypothesis should be interpreted with caution because our study included only patients with medication-resistant AVH, which was not the case in previous studies. The direction of structural connectivity changes in the interhemispheric auditory pathway needs to be better understood in SZ patients with AVH. These changes could depend on individual factors such as age, illness duration, but also antipsychotic treatment and severity of symptomatology.

**Correlations between white matter integrity and AVH severity**

We reported significant positive correlations between white matter integrity in the rostral portion of the direct segment of the left arcuate fasciculus and the attentional salience of AVH (i.e., the degree to which AVH capture attention and alter ongoing thought and behaviour). These results suggest that in SZ with AVH, the left arcuate fasciculus might be involved in information transfer across the language network and more specifically of its modulation by exogenous involuntary attention. A significant correlation between FA and the severity of AVH in SZ patients has been pinpointed in previous studies with regards to the whole left arcuate fasciculus (Rotarska-Jagiela et al., 2009; Ćurčić-Blake et al., 2015; Psomiades et al., 2016) or for the posterior genu of the corpus callosum (Knöchel et al., 2012). Our findings bring additional anatomical details to prior models and suggest for the first time precise white matter locations whose microstructure might be important for the generation of AVH. At difference with other work, we failed to find any significant association between attentional capture and any of the explored white matter pathways of the right hemisphere (Corbetta and Shulman, 2002).

Moreover, significant correlation between local FA value in the left arcuate fasciculus and distress associated to the AVH experience is difficult result to interpret and could be related to attentional salience, under the assumption that the higher the attentional salience, the higher the distress caused by AVH. In any case, the potential role of the arcuate fasciculus in the attentional capture process tied to AVH remains a very intriguing possibility in need of further study.

**Limitations**

Several limitations may impact the present work. First, our results which are consistent with previous published evidence, are based on a limited sample size, although we cannot completely rule out the possibility of a false positive caused by the reduced sample size. The findings of our study must be considered preliminary and must be corroborated in a larger sample of patients. Second, decreased white matter integrity could be induced artificially in areas of crossing fibres between the left arcuate fasciculus and rostral transcallosal fibres. Although we attempted to limit this possibility by using the spherical deconvolution method, a 45º crossing may affect the quality of fibre orientation. The use of a b-value of 2000 s/mm² during MRI acquisitions could have minimized uncertainty with regards to fibre crossings (Tournier et al., 2008). Nonetheless, to maximize signal-to-noise ratio, we opted for a b-value of 1500 s/mm² which is a widely accepted standard in the diffusion imaging field. Of note, the acquisition parameters, which often generate comparison biases, were identical for both groups of our study. Third, the weak or lack of an association between FA in the left arcuate fasciculus and AVH severity could be explained by our limited patient sample size. Nonetheless, to limit the effect of this bias, our analytical approach used a well-established conservative method particularly suitable to isolate the focal properties of white matter tracts (Colby et al., 2012; Yeatman et al., 2012). Fourth, a tractography study using an along-tract approach to investigate white matter integrity in the arcuate fasciculus and the interhemispheric pathways in SZ patients with AVH compared to those without AVH is needed to specifically -and directly- address the influence of white matter microstructural alterations on AVH. Replication in a larger sample of patients including SZ patients with and without AVH will allow a better understanding on whether these local alterations are the consequence of AVH, of the diagnosis of SZ, or a combination of both.

**CONCLUSION**

Our study highlights for the first time site-specific white matter microstructural abnormalities in the left arcuate fasciculus and the interhemispheric transcallosal fibres in SZ patients with medication-resistant AVH. Site specific FA differences could be explained either by the reinforcement of aberrant connections or via the deterioration of healthy axonal fibres. These outcomes emphasize the value of studying site-specific white matter bundles divided in specific sections along a given fasciculus. The specific consequences of local white matter abnormalities particularly for AVH remain to be clarified in the future, an effort that will require the comparison of white matter integrity between SZ patients with vs. without AVH.

Finally, current findings may prove useful to model and implement more efficient brain stimulation therapeutic strategies, such as those using rTMS or transcranial current stimulation (tCS) for AVH in SZ. These technologies aim at modulating network excitability patterns by spreading its effects via white matter connections. A better understanding of white matter integrity throughout the language network will provide a better understanding of the physiology subtending the effects of stimulation on AVH, and on such basis optimize neuromodulation therapeutic schemes or develop effective patient-customized strategies.

**Table 1.** Characteristics of the study samples

| Characteristics | SZ patients (n = 17) | Healthy subjects (n = 22) | *p* |
|---|---|---|---|
| **SOCIODEMOGRAPHIC** | | | |
| Age (years) | 41.47 (8.83) | 37.68 (11.12) | 0.257 |
| Sex n (% men) | 10 (58.82) | 11 (50) | 0.823 |
| Education (years) | 13.29 (3.46) | not acquired | - |
| Illness (years) | 12.82 (8.06) | - | - |
| **TREATMENT** | | | |
| CPZ eq (mg/day) | 502.79 (332.09) | - | - |
| typical AP(n) | 2 | - | - |
| atypical AP(n) | 11 | - | - |
| typical + atypical AP (n) | 4 | - | - |
| Clozapine (n) | 2 | - | - |
| **CLINICAL** | | | |
| AHRS | 24.18 (7.06) | - | - |
| *Frequency* | 4.41 (3.43) | - | - |
| *Reality* | 3.59 (1.37) | - | - |
| *Loudness* | 2.59 (1.12) | - | - |
| *Number of voices* | 2.29 (1.76) | - | - |
| *Length* | 3.24 (1.03) | - | - |
| *Attentional salience* | 4.53 (1.66) | - | - |
| *Distress level* | 3.53 (1.01) | - | - |
| PANSS | | | |
| *Positive* | 16.41 (4.65) | - | - |
| *AH (P3 item)* | 4.35 (1.22) | - | - |
| *Negative* | 19.53 (7.46) | - | - |
| *General* | 35.94 (7.30) | - | - |
| *Total* | 71.88 (16.42) | - | - |

Data are mean (SD). AP: Antipsychotic; CPZ eq: Chlorpromazine equivalents. AHRS: Auditory Hallucinations Rating Scale; HA: Hallucination; PANSS: Positive And Negative Syndrome Scale; PSYRATS HA: Psychotic Symptom Rating Scale.

**Table 2.** Correlation analyses between FA values along the left arcuate fasciculus, rostral corpus callosum and interhemispheric auditory pathway and the AHRS score

| Point along tract | Left arcuate fasciculus | | Rostral corpus callosum | | Interhemispheric auditory pathway | |
|---|---|---|---|---|---|---|
| | p-value | correlation coef. | p-value | correlation coef. | p-value | correlation coef. |
| 1 | 0.663 | 0.32 | 0.999 | -0.01 | 1 | -0.3394 |
| 2 | 0.976 | 0.05 | 1 | -0.24 | 1 | -0.1398 |
| 3 | 0.993 | -0.01 | 1 | -0.19 | 1 | -0.2095 |
| 4 | 0.996 | -0.04 | 1 | -0.24 | 1 | -0.1453 |
| 5 | 0.999 | -0.12 | 1 | -0.33 | 1 | -0.5721 |
| 6 | 1 | -0.22 | 1 | -0.09 | 1 | -0.3028 |
| 7 | 1 | -0.25 | 0.994 | 0.06 | 1 | -0.066 |
| 8 | 1 | -0.22 | 0.969 | 0.13 | 0.9836 | 0.1353 |
| 9 | 1 | -0.12 | 0.984 | 0.10 | 0.9946 | 0.0824 |
| 10 | 0.998 | -0.06 | 1 | -0.04 | 0.9984 | 0.0207 |
| 11 | 0.998 | -0.08 | 0.997 | 0.03 | 0.9986 | 0.0161 |
| 12 | 0.999 | -0.11 | 1 | 0.02 | 0.9976 | 0.0413 |
| 13 | 0.997 | -0.05 | 1 | -0.22 | 1 | -0.2048 |
| 14 | 0.950 | 0.11 | 1 | -0.34 | 1 | -0.2417 |
| 15 | 0.692 | 0.30 | 1 | -0.14 | 0.9864 | 0.1244 |
| 16 | 0.458 | 0.41 | 1 | -0.06 | 0.9992 | -0.0086 |
| 17 | 0.560 | 0.36 | 1 | -0.06 | 0.9712 | 0.1731 |
| 18 | 0.645 | 0.33 | 1 | -0.06 | 0.998 | 0.0319 |
| 19 | 0.705 | 0.30 | 0.895 | 0.23 | 0.9956 | 0.0694 |
| 20 | 0.746 | 0.28 | 0.661 | 0.35 | 1 | -0.152 |
| 21 | 0.752 | 0.28 | 0.869 | 0.25 | 0.9958 | 0.0676 |
| 22 | 0.755 | 0.27 | 0.783 | 0.30 | 0.999 | -0.003 |
| 23 | 0.769 | 0.27 | 0.794 | 0.29 | 0.998 | 0.0277 |
| 24 | 0.705 | 0.30 | 0.640 | 0.36 | 1 | -0.1421 |
| 25 | 0.678 | 0.31 | 0.796 | 0.29 | 1 | -0.3045 |
| 26 | 0.731 | 0.29 | 0.872 | 0.24 | 1 | -0.0949 |
| 27 | 0.842 | 0.22 | 0.991 | 0.07 | 1 | -0.1764 |
| 28 | 0.895 | 0.17 | 0.991 | 0.07 | 0.9996 | -0.0353 |
| 29 | 0.876 | 0.19 | 0.987 | 0.09 | 0.9438 | 0.2294 |
| 30 | 0.720 | 0.29 | 1 | -0.24 | 0.8706 | 0.3069 |
| 31 | 0.368 | 0.44 | 1 | -0.20 | 0.8364 | 0.3319 |
| 32 | 0.127 | 0.57 | 1 | -0.09 | 0.6728 | 0.418 |
| 33 | 0.142 | 0.56 | 1 | -0.36 | 0.6596 | 0.4234 |
| 34 | 0.797 | 0.25 | 1 | -0.29 | 0.9654 | 0.1887 |
| 35 | 0.992 | -0.01 | 1 | -0.11 | 0.8806 | 0.299 |
| 36 | 0.998 | -0.07 | 1 | -0.07 | 0.9914 | 0.1007 |
| 37 | 0.960 | 0.09 | 1 | -0.07 | 1 | -0.1705 |
| 38 | 0.889 | 0.18 | 1 | -0.23 | 1 | -0.2691 |
| 39 | 0.889 | 0.18 | 1 | -0.16 | 1 | -0.3282 |
| 40 | 0.949 | 0.11 | 1 | -0.13 | 1 | -0.4192 |
| 41 | 0.947 | 0.11 | 1 | -0.21 | 1 | -0.4037 |
| 42 | 0.670 | 0.32 | 1 | -0.08 | 1 | -0.1628 |

Notes: *p*-values are adjusted values after permutation tests (n = 5000). Coef.: coefficient

**Table 3.** Correlation analysis between local FA values of white matter pathways and clinical scores

|  | AHRS | | AHRS *attentional salience* | | AHRS *distress* | | PANSS | |
|---|---|---|---|---|---|---|---|---|
|  | **Coefficient r** | ***p*-value** | **Coefficient r** | ***p*-value** | **Coefficient r** | ***p*-value** | **Coefficient r** | ***p*-value** |
| **Left arcuate fasciculus** | | | | | | | | |
| *point 31* | 0.44 | 0.075 | 0.59 | **0.013** | 0.42 | 0.090 | -0.19 | 0.472 |
| *point 32* | 0.57 | 0.017 | 0.63 | **0.007** | 0.54 | 0.025 | -0.05 | 0.849 |
| *point 33* | 0.56 | 0.020 | 0.57 | 0.016 | 0.65 | **0.005** | 0.10 | 0.711 |
| *point 34* | 0.25 | 0.330 | 0.41 | 0.102 | 0.58 | 0.014 | 0.15 | 0.573 |
| **Corpus callosum (rostral portion)** | | | | | | | | |
| *point 8* | 0.13 | 0.606 | 0.55 | 0.022 | 0.66 | **0.004** | 0.24 | 0.354 |
| *point 9* | 0.10 | 0.697 | 0.39 | 0.125 | 0.68 | **0.003** | 0.18 | 0.493 |
| *point 10* | -0.03 | 0.890 | 0.01 | 0.961 | 0.33 | 0.200 | 0.03 | 0.906 |
| *point 18* | -0.06 | 0.820 | -0.30 | 0.235 | -0.18 | 0.490 | -0.28 | 0.272 |
| *point 25* | 0.29 | 0.257 | -0.09 | 0.740 | -0.07 | 0.797 | -0.22 | 0.395 |
| *point 26* | 0.24 | 0.345 | -0.24 | 0.364 | -0.06 | 0.811 | -0.09 | 0.743 |
| *point 27* | 0.07 | 0.779 | -0.37 | 0.139 | -0.16 | 0.524 | -0.14 | 0.591 |
| *point 34* | -0.29 | 0.267 | 0.20 | 0.438 | 0.27 | 0.287 | 0.08 | 0.768 |
| *point 35* | -0.11 | 0.685 | 0.26 | 0.318 | 0.50 | 0.041 | 0.11 | 0.680 |
| *point 36* | -0.07 | 0.786 | 0.27 | 0.296 | 0.53 | 0.029 | 0.02 | 0.951 |
| **Interhemispheric auditory pathway** | | | | | | | | |
| *point 17* | 0.17 | 0.554 | -0.09 | 0.753 | -0.26 | 0.374 | 0.28 | 0.341 |
| *point 18* | 0.03 | 0.914 | -0.21 | 0.481 | -0.46 | 0.099 | 0.34 | 0.229 |
| *point 19* | 0.07 | 0.814 | -0.02 | 0.944 | -0.36 | 0.202 | 0.06 | 0.849 |
| *point 20* | -0.15 | 0.604 | -0.33 | 0.252 | -0.47 | 0.093 | 0.04 | 0.902 |
| *point 21* | 0.07 | 0.818 | -0.16 | 0.584 | -0.22 | 0.445 | -0.20 | 0.501 |
| *point 22* | -0.00 | 0.992 | -0.25 | 0.396 | -0.54 | 0.047 | -0.09 | 0.747 |
| *point 23* | 0.03 | 0.925 | -0.24 | 0.401 | -0.39 | 0.172 | -0.03 | 0.926 |

Notes: *p*-values in bold indicated significant p-values after Bonferroni correction (p ≤ 0.0125). Pearson correlation coefficients (r) and *p*-values comparing SZ patients and healthy individuals are presented only for tract sites showing significant differences of FA.

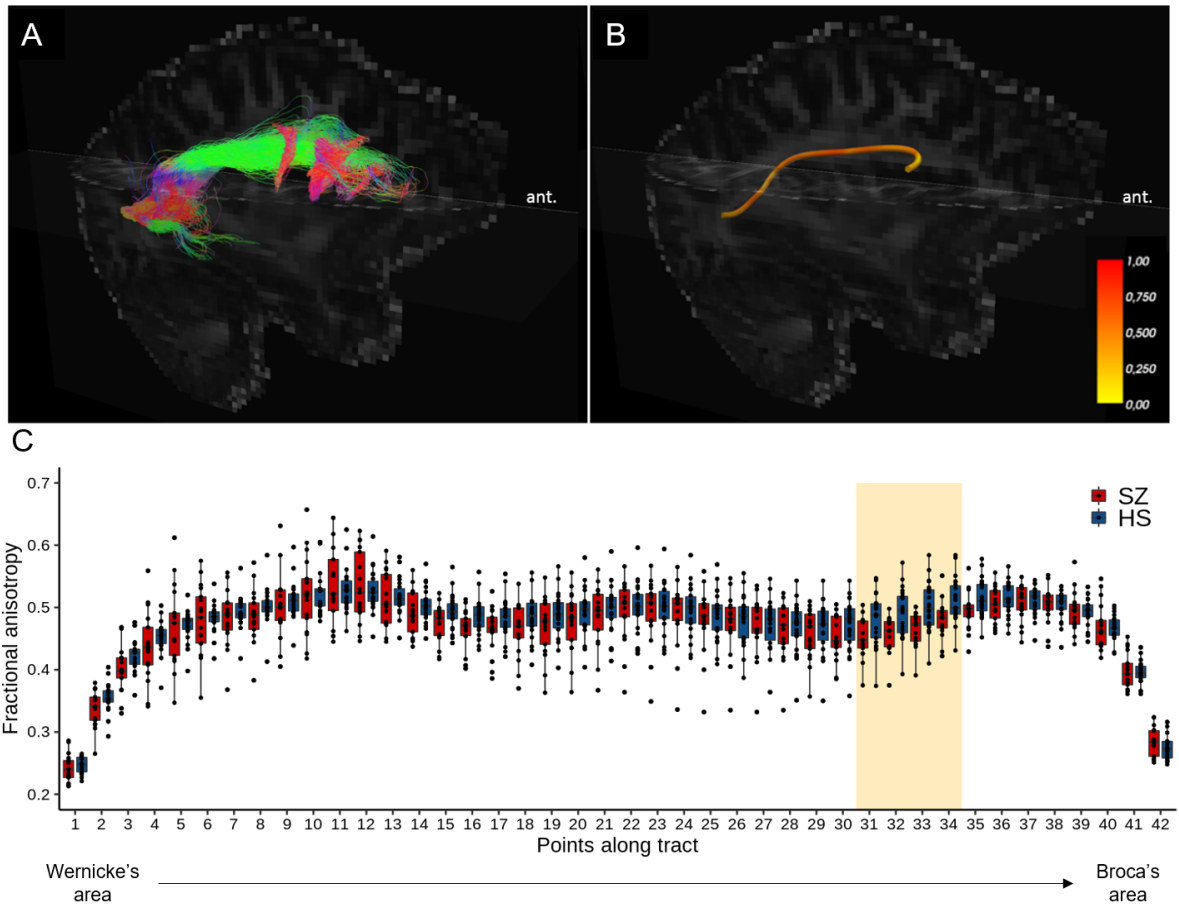

**Figure 1.** Tractographic reconstruction of the left arcuate fasciculus in schizophrenia (SZ) patients compared to healthy subjects (HS). (A) Reconstruction of the white matter fibres of the left arcuate fasciculus and (B) mean tract from a representative SZ patient, displayed in a sagittal view of the left hemisphere and superimposed on the MRI diffusion volume. Colour scale in the panel A represents the direction of the reconstructed streamlines (green: antero-posterior axis; red: left-right axis; blue: rostro-caudal axis). The colour bar in the panel B indicates fractional anisotropy (FA) values, from 0 to 1. (C) Boxplots represent the FA values measured in 42 contiguous points along the tract in patients (red boxplots) or control subjects (blue boxplots) from caudal to rostral. Point number 1 and 42 corresponds to the left posterior superior temporal gyrus and to Broca's area, respectively. The yellow box indicates white matter tract segments showing statistically significant FA differences between patients and control individuals. Ant.: anterior.

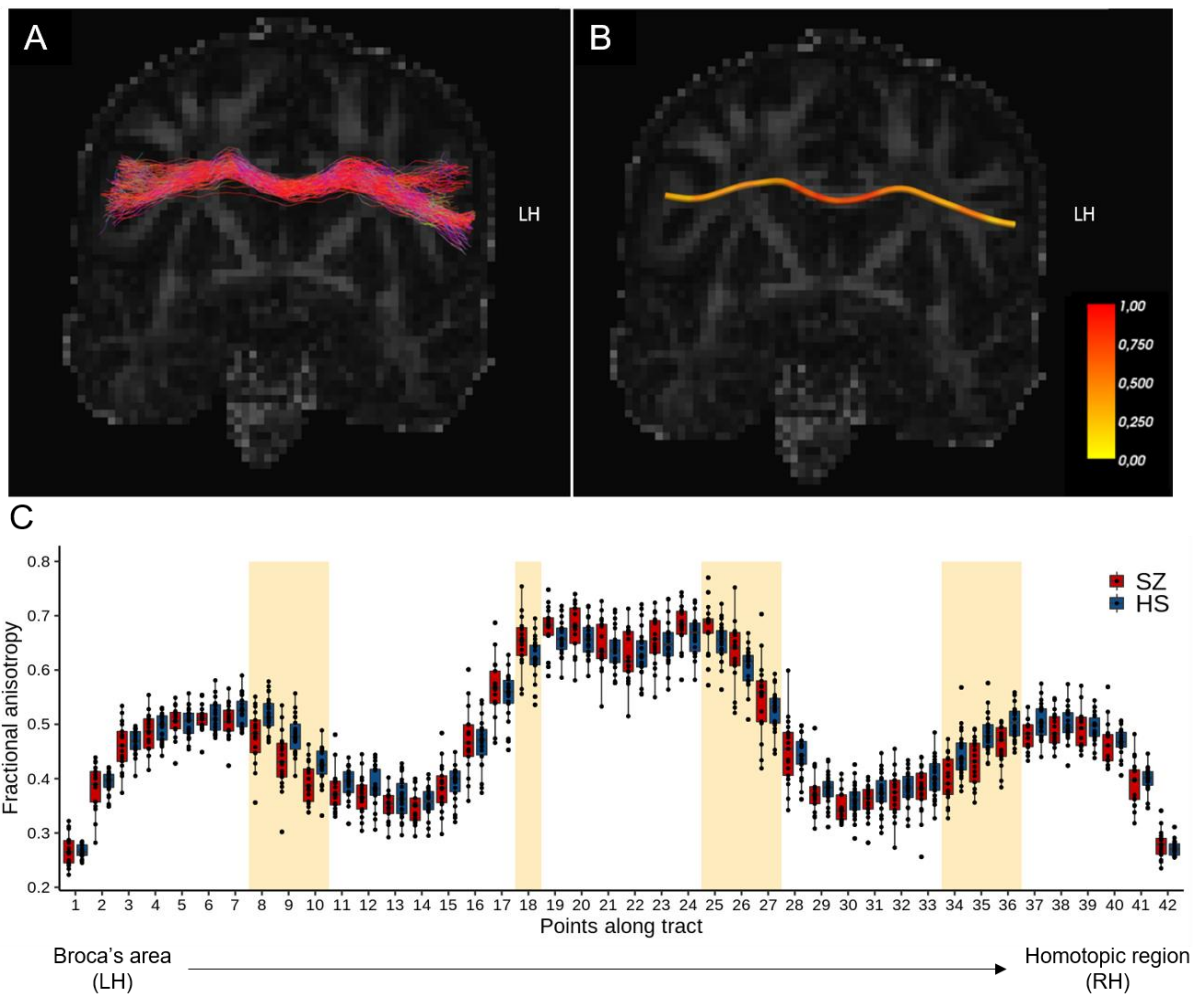

**Figure 2.** Integrity of interhemispheric fibres throughout the genu of the corpus callosum in schizophrenia (SZ) patients compared to healthy subjects (HS). (A) Reconstruction of the white matter fibres of the corpus callosum connecting Broca's area and its homotopic region in the right hemisphere and (B) mean tract from a representative patient displayed in a coronal MRI view and superimposed on the diffusion volume acquired during the diffusion-weighted MRI acquisition. The colour scale in the panel A represents the direction of the reconstructed streamlines (green: antero-posterior axis; red: left-right axis; blue: rostro-caudal axis). The colour bar in the panel B indicates the range of fractional anisotropy (FA) values, from 0 to 1. (C) Boxplots represent the FA values measured in 42 contiguous points along the tract in patients (red boxplots) and control subjects (blue boxplots). Point no.1 (left) corresponds to Broca's area whereas no. 42 (right) corresponds to the homotopic region of Broca's area in the right hemisphere. (C) The yellow box signals tract segments showing statistically significant FA differences between patients and control individuals. LH: left hemisphere.

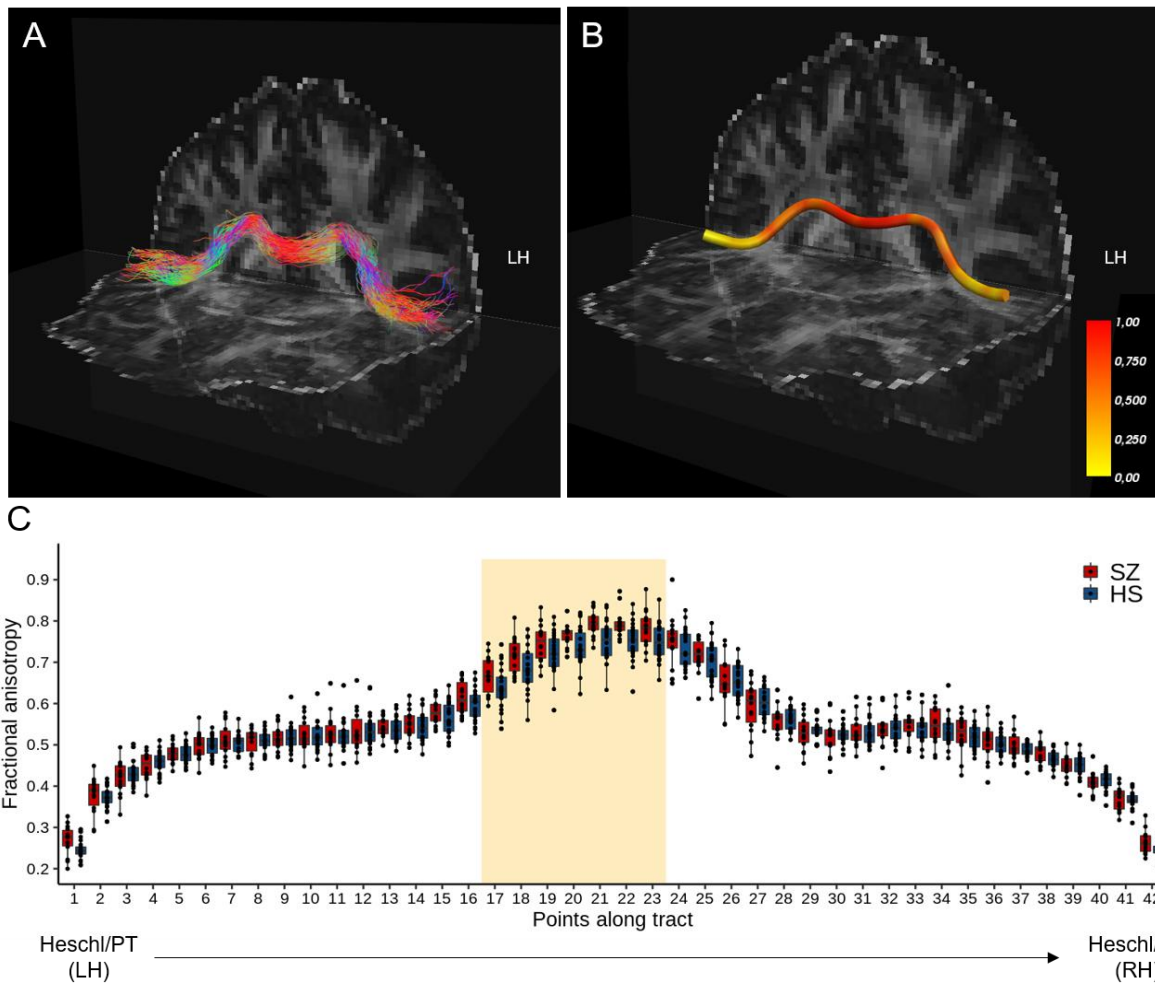

**Figure 3.** Integrity of interhemispheric fibres throughout the splenium of the corpus callosum in schizophrenia (SZ) patients compared to healthy subjects (HS). (A) Reconstruction of the white matter fibres of the corpus callosum connecting the left Heschl's gyrus/planum temporale (PT) to its homologous regions in the right hemisphere and (B) mean tract from a representative patient displayed in a coronal MRI view and superimposed on the diffusion volume acquired during the diffusion-weighted MRI acquisition. The colour scale in the panel A represents the direction of the reconstructed streamlines (green: antero-posterior axis; red: left-right axis; blue: rostro-caudal axis). The colour bar in the panel B indicates the range of fractional anisotropy (FA) values, from 0 to 1. (C) Boxplots represent the FA values measured in 42 contiguous points along the tract in patients (red boxplots) and control subjects (blue boxplots). Point no.1 (left) corresponds to the left Heschl's gyrus/planum temporale whereas no. 42 (right) corresponds to the right Heschl's gyrus/planum temporale. (C) The yellow box signals tract segments showing statistically significant FA differences between patients and control individuals. LH: left hemisphere.

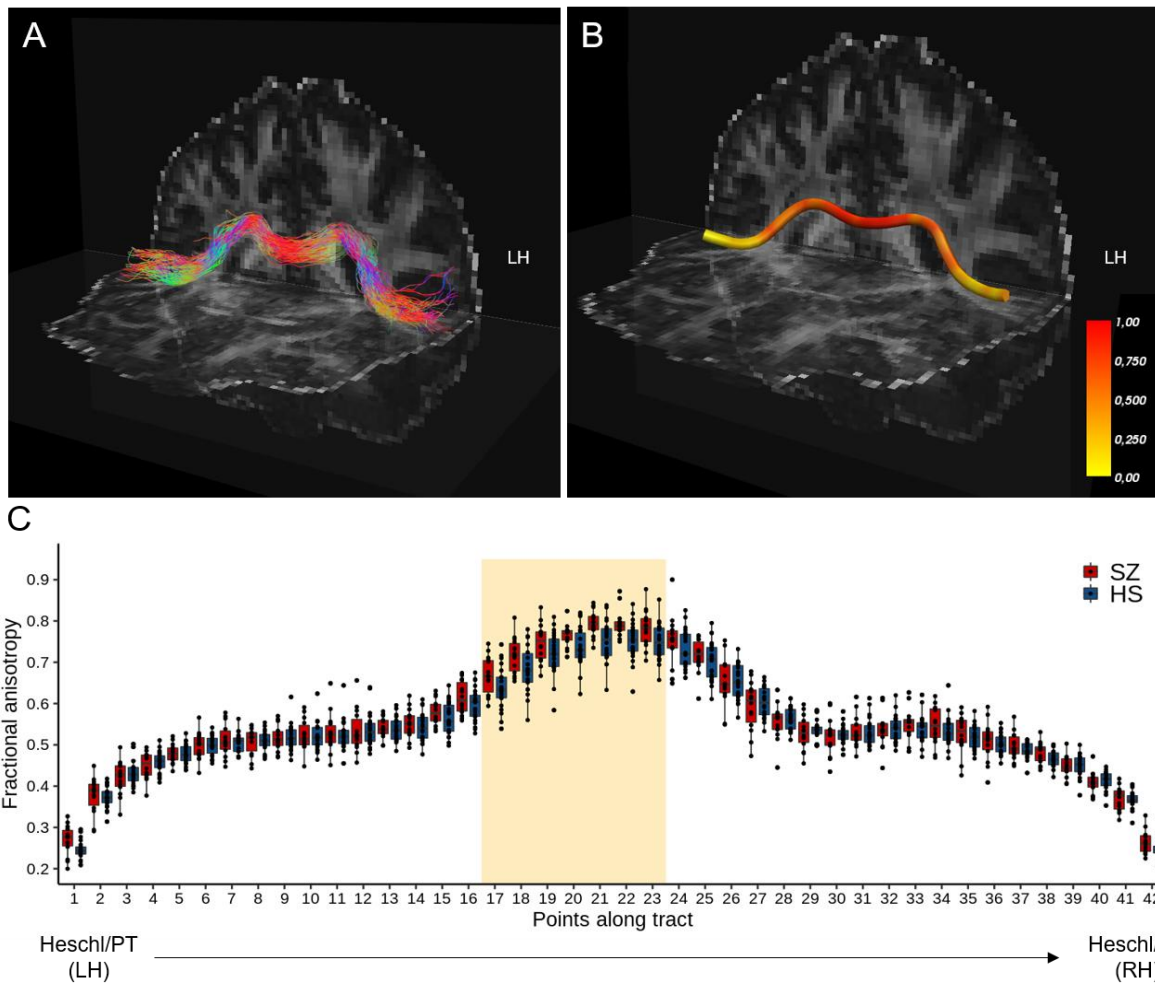

**Figure 3.** Integrity of interhemispheric fibres throughout the splenium of the corpus callosum in schizophrenia (SZ) patients compared to healthy subjects (HS). (A) Reconstruction of the white matter fibres of the corpus callosum connecting the left Heschl's gyrus/planum temporale (PT) to its homologous regions in the right hemisphere and (B) mean tract from a representative patient displayed in a coronal MRI view and superimposed on the diffusion volume acquired during the diffusion-weighted MRI acquisition. The colour scale in the panel A represents the direction of the reconstructed streamlines (green: antero-posterior axis; red: left-right axis; blue: rostro-caudal axis). The colour bar in the panel B indicates the range of fractional anisotropy (FA) values, from 0 to 1. (C) Boxplots represent the FA values measured in 42 contiguous points along the tract in patients (red boxplots) and control subjects (blue boxplots). Point no.1 (left) corresponds to the left Heschl's gyrus/planum temporale whereas no. 42 (right) corresponds to the right Heschl's gyrus/planum temporale. (C) The yellow box signals tract segments showing statistically significant FA differences between patients and control individuals. LH: left hemisphere.